\pdfoutput=1
\documentclass[sigconf, nonacm]{acmart}

\usepackage{caption}
\usepackage{subcaption}

\AtBeginDocument{%
  \providecommand\BibTeX{{%
    \normalfont B\kern-0.5em{\scshape i\kern-0.25em b}\kern-0.8em\TeX}}}

\usepackage{amsmath,amsfonts}
\usepackage{algorithmic}
\usepackage{graphicx}
\usepackage{textcomp}
\usepackage{xcolor}
\usepackage{bigstrut} \usepackage{multirow} 
\usepackage{algorithm}
\usepackage{amsfonts}

\newcommand{\N}[0]{\mathcal{N}}

\begin{document}

\title{MiCRO: Multi-interest Candidate Retrieval Online}

\author{Frank Portman}
\email{fportman@twitter.com}
\affiliation{%
  \institution{Twitter Cortex}
  \city{Boston}
  \state{MA}
  \country{USA}
}

\author{Stephen Ragain}
\email{sragain@twitter.com}
\affiliation{%
  \institution{Twitter Cortex}
  \city{Chicago}
  \state{IL}
  \country{USA}
}

\author{Ahmed El-Kishky}
\email{aelkishky@twitter.com}
\affiliation{%
  \institution{Twitter Cortex}
  \city{Seattle}
  \state{WA}
  \country{USA}
}
\renewcommand{\shortauthors}{Portman, et al.}

\begin{abstract}
Providing personalized recommendations in an environment where items exhibit ephemerality and temporal relevancy (e.g. in social media) presents a few unique challenges: (1) inductively understanding ephemeral appeal for items in a setting where new items are created frequently, (2) adapting to trends within engagement patterns where items may undergo temporal shifts in relevance, (3) accurately modeling user preferences over this item space where users may express multiple interests. In this work we introduce MiCRO, a generative statistical framework that models multi-interest user preferences and temporal multi-interest item representations. Our framework is specifically formulated to adapt to both new items and temporal patterns of engagement. MiCRO demonstrates strong empirical performance on candidate retrieval experiments performed on two large scale user-item datasets: (1) an open-source temporal dataset of (\emph{User}, \emph{User}) \emph {follow} interactions and (2) a temporal dataset of (\emph{User}, \emph{Tweet}) \emph{favorite} interactions which we will open-source as an additional contribution to the community.

\end{abstract}

\settopmatter{printacmref=false}
\setcopyright{none}
\renewcommand\footnotetextcopyrightpermission[1]{}

\maketitle

\section{Introduction}
\label{sec:intro}
Recommender systems are an important component of many web applications such as e-commerce and social media. These systems aim to provide users with relevant content in the form of ranked lists of items~\cite{aggarwal2016recommender}. The crux of the recommendation problem is to rank sets of items based on their relevance to each user. Since industrial recommender systems operating at web scale may have a large corpus of potential items to consider, the standard approach is to decompose this problem into two steps: (1) a lightweight candidate retrieval step that retrieves a high recall set of items and (2) a heavier ranking step to further prune and reorder the retrieved candidates~\cite{kang2019candidate}. It is important for the candidate retrieval step to return as many relevant items as possible to ensure a high-recall set for the ranker to optimize. If the users have diverse interests then the candidates should also be diverse in order to maximize recall~\cite{el2022knn,pal2020pinnersage}.

With the proliferation of deep learning methods for recommender systems~\cite{covington2016deep,cheng2016wide,wang2017deep}, a standard strategy for candidate retrieval has been to embed users and items into the same vector space and then use Approximate Nearest Neighbor (ANN) retrieval to find relevant items when queried via learned user vectors \cite{malkov2014approximate}. However, this paradigm can present a few problems in certain settings. First, it has been shown that without learning multiple representations for users and items, retrieved items tend to be highly similar to each other (i.e. pertaining to a single modal ``interest'')~\cite{wilhelm2018practical,el2022knn,zhang2020towards}. Second, recommending relevant items when new items are rapidly created (e.g., social media posts) is challenging\textemdash many collaborative filtering embedding techniques are \textit{transductive} and thus unable to embed out-of-vocabulary (OOV) items. To quickly adapt to new-item appeal and recommend these items to the proper user audience, it is crucial to perform candidate retrieval without having to retrain transductive embedding models to include newly created items.

In our work, we address these challenges by modeling user preferences as mixtures over interests, and interests as temporal distributions over items. In order to estimate model parameters, we develop a collapsed Gibbs-sampling method to efficiently allocate both new and existing items to interests. Aside from an initialization step to learn a graph embedding from a bipartite (\emph{User}, \emph{Item}) graph, our approach does not require re-training or otherwise updating any underlying user or item embeddings. Instead, we use observed historical engagement to directly build an estimated mixture representation of user interests. We use this estimate to induct mixture distributions on existing or OOV items as they are engaged with. Finally, we use our inferred user and interest parameters to estimate user-item engagement probabilities, allowing us to  conduct model-based candidate retrieval for each user. 

We benchmark our model-based candidate retrieval against standard ANN methods and a global popularity baseline for two large empirical datasets, finding that it improves on standard information retrieval (IR) metrics on both tasks.

\section{Preliminaries}
\label{sec:prelim}

Our data can be expressed as a bipartite graph $\mathcal{G}$ representing the engagements between users ($\mathcal{U}$) and items ($\mathcal{I}$) where each edge is associated with an ordinal time chunk. For each user and item, we observe a binary `relevance' variable indicating the item's relevance to that particular user. An item is considered relevant to a particular user if the user engages with the item (e.g., click, purchase, follow, like)

\subsection*{\textbf{Problem Formulation}}
Given an input user-item engagement graph $\mathcal{G}$ and a parameter, $K$, denoting the number of latent user interests, our aim is to model future user-item engagements for the purpose of generating candidates at a scale typical for an industrial recommender system setting. Specifically, we focus on addressing the candidate retrieval problem in settings where new items are created frequently and items may undergo temporal shifts in relevance.

\subsubsection*{{\bf Desired Properties:}}
\begin{itemize}
\item User preferences should be over \textit{multiple interests}.
\item Model-based recommendation of out-of-vocabulary items should be possible.
\item Our approach should capture temporal trends in item relevance.
\item Our approach should perform well on ephemeral items (i.e., items that are only relevant for short periods of time).
\end{itemize}

In order to model multiple user interests, we seek to learn two types of latent distributions: (1) multinomial distributions over items for each interest reflecting the distribution of item engagements corresponding to that interest and (2) multinomial distributions over interests for each user reflective of each user's interest preferences. We will further extend interest multinomials to be time-dependent in order to capture temporal trends and quickly allocate engagements of OOV items to interests. 

More formally, we seek to learn parameters $\phi_{k, t}$ of a topically-coherent multinomial distribution over items corresponding to the $k$-th interest and temporally relevant for time chunk $t$. Here, $\phi_{k,t}(i) = p(i|k, t)\in [0,1]$ is the conditional probability that a user engagement is with item $i$ given they are engaging with an item in interest $k$ at time $t$. For example, when modeling user-Tweet recommendation on Twitter, in an interest cluster that is predominantly ``machine learning'', we would expect to find Tweets from machine learning researchers on topics such as ``deep learning'' or ``statistical learning'' with high probability; however we would likely expect lower probability for Tweets from ``political theory'' academics discussing politics. 

Additionally we seek to learn parameters $\theta_{u}$ of a multinomial distribution modeling the user's distribution over these interests such that $\theta_{u,k}=p(k|u)\in [0,1]$ is the probability that an engagement from user $u$ arises from interest $k$. 

We aim to infer $\phi_{k,t}$ and $\theta_u$ such that under the associated generative model, the distributions accurately model user preferences and item appeal as evidenced by higher recall in a candidate retrieval task.

To perform item induction and candidate retrieval in a way that satisfies our desired properties, we propose our Multi-interest Candidate Retrieval Online (MiCRO) framework. MiCRO can be summarized as a four-step framework:
\begin{enumerate}
    \item A graph embedding and clustering approach to learn initial user preferences and item clusters.
    \item A temporal generative model allowing for temporal and ephemeral trends in item appeal and relevance to interests.
    \item A Gibbs-sampling based inference method for updating user and interests parameters as novel data arrives. 
    \item Model-based candidate generation.
\end{enumerate}

An additional benefit of our approach is that it allows us to incorporate recent engagements to update our estimation of user interests. Although we do not directly require an updating of user interests using OOV items as part of our desired properties, this provides practical benefits and can improve performance.

\subsection{Multi-interest Modeling}
\label{sec:graph}

In this section we consider a graph-embedding and clustering approach to model users over interests and interests over items. While this initial multi-interest retrieval model can be promising in certain settings, it is transductive and thus cannot handle temporal/ephemeral item appeal or new OOV items. We later address these shortcomings in our proposed framework, MiCRO, while making use of these  user preferences over embedding-based interest clusters as a way to initialize MiCRO.

\subsubsection{\bf{Co-embedding  Users and Items}} \label{making_an_embedding}
While our approach is agnostic to the method used to co-embed users and items, and many off-the shelf approaches can be utilized without loss of generality, we outline a simple and scalable approach to co-embedding users and items for completeness. Taking inspiration from the approach outlined in~\cite{el2022twhin,el2022knn}, we create a bipartite graph ($\mathcal{G}$) between users ($\mathcal{U}$) and items ($\mathcal{I}$) where an edge represents a user-item engagement and is associated with some ordinal time chunk.

We seek to learn shallow embedding vectors (i.e., vectors of learnable parameters) for each user ($u_j$) and item ($i_k$) in this bipartite graph; we denote these learnable embeddings for users and items as $\mathbf{u_j}$ and $\mathbf{i_k}$
respectively. A user-item pair is scored with a scoring function of the form $f(\mathbf{u_j}, \mathbf{i_k})$. Our training objective seeks to learn $\mathbf{u}$  and $\mathbf{i}$ parameters that maximize a log-likelihood constructed from the scoring function for $(u, i) \in \mathcal{G}$ and minimize for $(u, i) \notin \mathcal{G}$. Embedding vectors can then be learned using turn-key knowledge graph embedding techniques such as TransE~\cite{bordes2013translating}.

\subsubsection{\bf{Learning User Preferences and Item Clusters}} \label{clustering_and_projecting}
As previous works~\cite{el2022knn, pal2020pinnersage} have shown, clustering item representations can partition the item space into topically coherent interest clusters representative of diverse user interests. These clusters can be used to build multinomial distributions that capture user preferences over interests. To do this, we first perform spherical $k$-means clustering~\cite{Dhillon2004ConceptDF} in the item space to group topically similar items into $k$ \textit{interest clusters}. 

Given these interest clusters, we can write the full distribution $p(i | u)$ as a mixture over interest clusters $ p(i | u) = \sum_k p(k | u) \cdot p(i | k)$. Using the item clusters, we get straightforward counting-based Maximum Likelihood estimators (MLEs) for the user-interest engagment probabilities. Namely, $
p_{\mathrm{mle}}(k|u)$ is proportional to the number of times $u$ has engaged with an item in cluster $k$. Similarly we can compute the MLE estimates of $p_{\mathrm{mle}}(i|k)$ as proportional to the number of engagements with item $i$, normalized by the total number of engagements with cluster $k$. 

Summarizing this, we are modeling each user's higher level interests, $p(k | u)$, and then within each interest $k$, we are modeling a distribution over items, $p(i | k)$, by considering all engagements to the item $i$ as ``belonging'' to the interest $k$ that it was clustered into. Plugging $p_{\mathrm{mle}}(i|k)$ and $p_{\mathrm{mle}}(k|u)$ into $p(i | u) = \sum_k p(k | u) \cdot p(i | k)$ yields $p_{\mathrm{mle}}(i | u)$.

\section{M\lowercase{i}CRO Framework}
\label{sec:generative}
In Section~\ref{sec:graph}, we described a graph embedding and clustering based method to represent users and interests as static mixture distributions. Because such a model is transductive and fails in the trending and ephemeral item setting, we instead introduce a temporal generative model. Our generative model for a directed temporal bipartite graph consists of a set of users $\mathcal{U}$ that create engagements to a set of items $\mathcal{I}$ over a series of time chunks $1,\dots,T$.

Similar to LDA~\cite{blei2003latent}, we propose a generative model based on a ``bag-of-items'' assumption where users engage with items via latent vectors of users and interests. We seek to associate each engagement with one of $K$ interests. Each of these $K$ interests has a latent distribution $\phi_{k,t}$ which represents the distribution over items in interest $k$ at time $t$. We use a prior $Dir(\beta)$ for this latent distribution. 

Each user $u \in U$ has a latent distribution $\theta_u$ over these interests representing the affinity that a user has for each of the $K$ interests. While we use the same prior $Dir(\beta)$ for every $\phi_{k,t}$, our priors for $\theta_u$ are both user-dependent and more constrained. Namely, we use $Dir(\alpha_u)$ priors for $\theta_u$, where we typically model $\alpha_u$ as sparse, as each user typically engages with a small proportion of the $K$ total interests. 

Given these parameters, we can now describe the generative process for engagements. During time window $t$, each user $u$ engages with each of $N_{u,t}$ items. For the $j$-th of these items, the the user first selects an interest $z_{u,j} \in \{1,\dots,K\}$ from  $Multi(\theta_u)$. Next the user selects an item $i_{u,m}$ to engage with according to $Multi(\phi_{z_{u,j},t})$. The complete generative process is detailed in Table~\ref{tab:generative_story}. %

Because all $\theta_u$ and $\phi_{k,t}$ are unobserved, we use a Gibbs sampler to estimate the latent engagement interests and update our parameter estimates. When sampling, we process one time chunk at a time, starting with aggregates from the previous time chunk. When moving to the next time chunk, we treat our final latent engagement interest samples as observed. 
\\ \

\begin{table}[h!]
\centering
\vspace{-0.5cm}
\vspace{0.05cm}
      \caption{Multi-interest Candidate Retrieval Model Notation}
    \begin{tabular}{|p{1.2cm}|p{6.9cm}|}
    \hline
    \textbf{Variable} & \textbf{Description} \bigstrut\\
    \hline
    $U$, $K$, $I$ & number of users, interests, items \bigstrut \\
    \hline
    $u$, $k$, $i$  & user index, interest, item \bigstrut \\
    \hline
    $\theta_{u}$ & user $u$'s multinomial interest distribution \bigstrut \\
    \hline
    $i_{u,j}$ & the $j_{th}$ item engaged with by user $u$ \bigstrut \\
    \hline
    $t_{u,j}$ & the time index when engagement $i_{u,j}$ occurred  \bigstrut \\
    \hline
    $z_{u,j}$ & latent interest corresponding to engagement $i_{u,j}$ \bigstrut \\
    \hline 
    $e_t$ & set of engagements during time $t$, i.e. $e_t = \{ i_{u,j} \in E : t_{u,j} = t\}$ \bigstrut \\ 
    \hline
    $E_t$ & set of all engagements created up to time $t$, i.e. $E_t = \{ i_{u,j} \in E : t_{u,j} \leq t\}$ \\
    \hline 
    $z_t$ & latent interests for all engagements at time $t$, i.e. $z_t = \{z_{u,j} : i_{u,j} \in e_t\}$ \\
    \hline
    $Z_t$ & latent interests for all engagements up to time $t$, i.e. $Z_t = \{z_{u,j} : i_{u,j} \in E_t\}$ \\
    \hline 
    $\phi_{k, t}$ & multinomial distribution over items in engagements of interest $k$ for time $t$ \bigstrut \\
    \hline
    $N_{u,t}$ & number of engagements made by user $u$ in time $t$ \bigstrut \\
    \hline
    $\mathcal{N}_{k,t}$ & $\mathcal{N}_{k,t} = \sum_{i_{u,j} \in e_t} \mathbf{I}(z_{u,j} = k)$, number of engagements in interest $k$ during time period $t$\bigstrut \\
    \hline
    $\mathcal{N}_{u,k,t}$ & $\mathcal{N}_{u,k,t} = \sum_{ i_{u',j} \in E_t} \mathbf{I}(u' = u, z_{u',j} = k)$, number of engagements with interest $k$ for $u$ up to time $t$ \bigstrut \\
    \hline
    $\mathcal{N}_{i,k,t}$ & $\mathcal{N}_{i,k,t} = \sum_{i'_{u,j} \in e_t} \mathbf{I}(i' = i, z_{u,j} = k)$, number of engagements to item $i$ during time period $t$ that were affiliated with interest $k$ \\
    \hline
    $\alpha_u, \beta$ & prior parameters of the Dirichlet distributions for $\theta_u,\phi_{k,t}$ respectively \bigstrut \\
    \hline
    \end{tabular}%
  \label{tab:DLMmodel}%
\end{table}%

\begin{table}[ht]
\centering
\normalsize
\vspace{0.05cm}
\caption{MiCRO Generative Model}
\begin{tabular}{|l|}
 \hline
For user $u = 1,2,\dots,U$: \\
\quad Draw $\theta_u \sim Dir(\alpha_u)$ \\
For time period $t=1,2,\dots,T$:\\
\quad For $k = 1,2,\dots,K$: \\
\quad \quad Draw $\phi_{k,t} \sim Dir(\beta)$\\ 
\quad For each $u = 1,2,...,U$, and $j_\textit{th}$ item, $j = 1,2,...,N_{u,t}$:\\
\quad \quad \quad Draw $z_{u,j} \sim Multi(\theta_u)$ \\
\quad \quad \quad Draw $i_{u,j} \sim Multi(\phi_{z_{u,j},t})$ \\
\hline
\end{tabular}
\label{tab:generative_story}%
\end{table}

\begin{figure}
\includegraphics[width=8cm]{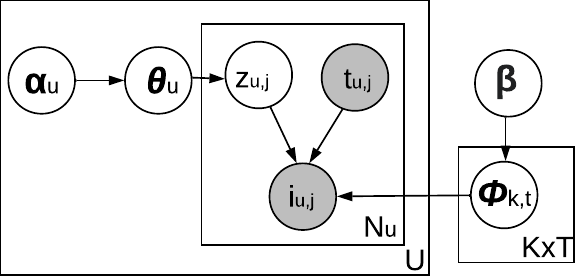}
\caption{In MiCRO, a latent interest $z_{u,j}$ is drawn from the user's interest distribution ($\theta_u$). The interest $z_{u,j}$ alongside the current observable time chunk $t_{u,j}$ is used to draw an item from the temporally relevant item distribution $\phi_{k,t}$.}
\label{fig:plate}
\end{figure}

\subsection{Model Inference}
The graphical model for our Multi-interest Candidate Retrieval Online (MiCRO) framework, depicted in Figure~\ref{fig:plate}, defines the joint distribution of random variables. By utilizing the conditional independence encoded in the graph, the joint distribution can be written as (we omit the hyper-parameters $\alpha~{and}~\beta$ for simplicity):

\begin{small}
\begin{align*}
\begin{split}
P(Z,E,T,\Phi,\Theta) = \prod_{u,j} p(z_{u,j}|\theta_u)p(i_{u,j}|z_{u,j},t_{u,j},\Phi) \prod_{u} p(\theta_u) \prod_{k,t}p(\Phi_{k,t})
\end{split}
\end{align*}
\end{small}
Where $E$ is the set of the entire user engagement history, $T$ is the set of discretized time chunks associated with each user-item engagement, and $Z$ is the set of all latent engagement interests. Notation for all derivations can be found in Table~\ref{tab:DLMmodel}. 

Because exact inference over the hidden interest variables within MiCRO is intractable due to the large number of hidden variables and parameters, we utilize collapsed Gibbs sampling to perform approximate inference. To reduce the uncertainty introduced by our multinomials, $\theta$ and $\phi$, we first integrate out these distributions by exploiting the conjugacy between the multinomial and Dirichlet distributions. This leaves a collapsed version of the joint probability distribution without any multinomial factors %
which we can write in closed form. For our joint distribution, we develop a collapsed Gibbs sampling algorithm to sample the latent assignment variables, $Z$, from its posterior. After integrating out $\Theta$ and $\Phi$ from the joint probability distribution, we can write our joint probability distribution as:

\begin{align*}
\begin{split}
P(Z_T,E_T) &\propto  \prod_{k=1}^{K} \prod_{u=1}^{U}\Gamma(\alpha_k+ \mathcal{N}_{u,k})  \prod_{t=1}^T \frac{\prod_{i=1}^{I} \Gamma(\beta_{t,i} + \mathcal{N}_{i,k, t})}{\Gamma(\sum_{i=1}^{I} \beta_{t,i} + \mathcal{N}_{k,t})}
\end{split}
\end{align*}

This derivation is analogous to the derivation from~\cite{griffiths2002gibbs}. Subsequently, the probability of a particular latent interest for an engagement on an item is computed conditional on all other latent assignments. For example, for engagement $i_{u,j}$ considering all engagements up through time $t$ for any time $t=1,\dots,T$, we can compute as follows:

\begin{align}
\label{eq:interest_link}
p(&z_{u,j} = k|E_t, Z_t \setminus z_{u,j})\; 
  \propto \;   \frac{ \Gamma(\alpha_{k}+\mathcal{N}_{u,k,t \setminus z_{u,j}} + 1)}
 { \Gamma(\alpha_{k}+\mathcal{N}_{u,k\setminus z_{u,j}}) }\times   \nonumber \\
&\hspace{-0.0cm}\frac{\Gamma(\beta_{t,i_{u,j}}+\mathcal{N}_{i,k,t\setminus z_{u,j}}+1 )}
{\Gamma(\sum_{i=1}^{I} \beta_{t,i}+\mathcal{N}_{k,t \setminus z_{u,j}}+1)}
/
\frac{\Gamma(\beta_{t,i_{u,j}} + \mathcal{N}_{i,k,t \setminus z_{u,j}} )}
{\Gamma (\sum_{i=1}^{I} \beta_{t,i} + \mathcal{N}_{k,t\setminus z_{u,j}} )} \;
 \nonumber \\
&=(\alpha_{k} + \mathcal{N}_{u,k,t\setminus z_{u,j}})
\frac{
( \beta_{t,i_{u,j}} + \mathcal{N}_{k,t,i\setminus z_{u,j}})
}{
(\sum_{i=1}^{I} \beta_{t,i}) +\mathcal{N}_{k,t\setminus z_{u,j}}
}
\end{align}

where we utilize the fact that $\Gamma(x+1)=x\Gamma(x)$. Note that as of time $t$, we've only observed the engagements up to but not including time $t$. 

Using this we have a straightforward Gibbs sampling approach for estimating the latent variables at time $t$ given all of the latent variables through time $t{-}1$. We start by initializing our estimate for the latent engagement interests for those engagements in time $t$ according to a distribution $D$. We then repeatedly loop over the engagements for this time period, resampling a new estimate for each engagement according to the Gibbs sampler and updating all of the corresponding counters appearing in the Gibbs distribution. We repeat this process until the chain has converged, and then move to the next time period, treating our estimates $z_t$ as observed going forward.

\begin{algorithm}
    \caption{Gibbs Sampler for MiCRO at time $t$}
\begin{algorithmic}
    \STATE Initialize $\hat{z_t} \sim D$
    \STATE Update $\hat {\mathcal{N}_{u,k,t}}, \hat {\mathcal{N}_{i,k}},  \hat {\mathcal{N}_{i,k,t}}$ given $\hat{z_t}$
    \WHILE{not converged}
        \FOR{$i_{u,j} \in e_t:$}
            \STATE Update $\hat z_{u,j} \sim p(z_{u,j} = k | E_t,\hat Z_{t,\setminus z_{u,j}})$
            \STATE Update $\hat {\mathcal{N}_{u,k,t}}, \hat {\mathcal{N}_{i,k}},  \hat {\mathcal{N}_{i,k,t}}$ given $\hat z_{u,j}$
        \ENDFOR
    \ENDWHILE
\end{algorithmic}
\end{algorithm}

Note that updating the counts $\hat {\mathcal{N}_{u,k,t}}, \hat {\mathcal{N}_{i,k}},  \hat {\mathcal{N}_{i,k,t}}$ given $\hat z_{u,j}$ is equivalent to updating our estimates of the user interest parameters $\theta_u$ and the interest item distribution $\phi_{k,t}$. We could write Equation~\ref{eq:interest_link} in terms of these estimates $\hat \theta_u$ and $\hat \phi_{k,t}$ as functions of our updated estimates of the latent engagement interests $\hat Z_t \setminus z_{u,j}$, but in practice we do not need to normalize these distributions and simply maintain the counts. Instead we can simply maintain current estimates of these counts as sufficient statistics of the latent interest probabilities as we run the Gibbs sampler. 

\subsection{Initialization}
In order to improve our parameter estimation, we initialize the user parameters ($\alpha_u$ and $\theta_u$) using aggregations from the graph embedding and clustering step described in Section~\ref{sec:graph}. We can think of this data as $t=0$ data or ``train'' data for notational convenience, while our $t=1,\dots,T$ estimations represent a ``test'' dataset where one might use MiCRO to perform online item induction and candidate retrieval. In this setting, we believe the clustering algorithm provides a good categorization of user engagement, but we are simply not able to re-embed and re-cluster the data quickly enough to pick up on emerging trends or embed novel items. Instead, we use the Gibbs sampler to propagate past embeddings through recent data by treating engagements to these clusters as observations of the latent user interests in the mixture model. 

In Section~\ref{sec:graph}, we compute a single interest (cluster) for each train set item. For our initialization, we associate every engagement with the item with the corresponding interest. Let $k_i$ be the cluster of item $i$ for each $i$ appearing in the train set.

Using our shorthand, we have $\hat{z_0} = \{k_i : (u,i) \in e_0\}$. 
Given $\hat{z_0}$, we get straightforward counting estimators for $\N_{u,0}, \N_{k,0}, \N_{u,k,0},$ and $\N_{i,k,0}$, which we can use to start all of the counters for the Gibbs sampler with values from our train data. We also set our user-level prior $\alpha_u$ to be a vector with value $\alpha$ at each observed interest, i.e. 
\[
(\alpha_u)(k) = \alpha \cdot \mathbf{1}(N_{u,k,0} > 0),
\]
where $\mathbf{1}(\cdot)$ denotes the indicator function. 

The intuition behind this user-specific prior is that users generally exhibit a few interests, and as such we can maintain a sparse user-specific prior resulting in significantly faster inference and superior empirical performance. We use these estimators based on the training (i.e. $t=0$ data) to initialize $\theta_u$
for each time period. In the notation of the Gibbs sampler probabilities as written in equation~\ref{eq:interest_link}, this is equivalent to using the underlying count estimates $\hat {\N_{u,0}}$ and $\hat{\N_{u,k,0}}$.
from time $t=0$ as our starting count estimates for each time $t=1,\dots,T$. For MiCRO, we choose $D$, the distribution from which $\hat{z_t}$ is initialized, to be $Dir(\alpha_u)$, i.e. the uniform prior for this $\theta_u$ obtained from the initial training data. We can think of setting the support for the prior using this initial training data as akin to a ``empirical Bayes" approach \cite{ragain2018improving, maritz2018empirical}.

We could instead consider similarly initializing $\phi_{k,t}$ using $\N_{k,0}$ and $\N_{i,k,0}$, but find that it offers no performance benefit in our data. This reinforces our decision to model $\phi_{k,t}$ as independent of $\phi_{k,t'}$ for earlier time chunks $t' < t$.

\subsection{Candidate Retrieval}
For time period $t$, once we have inferred user interest parameters $\theta_{u}$ and temporal item-interest parameters $\phi_{k,t}$ for all users and interests, we can estimate the most likely items for $u$ to engage with under the model, and return these as candidates.

Recalling that our generative model selects an interest $k$ via $\theta_{u}$ and then an item according to $\phi_{k,t}$, the probability that a given engagement is on item $i$ for user $u$ is can be expressed as a simple function of our model parameters. Letting $p_t(i | u)$ denote the probability that user $u$ engages with item $i$ in time chunk $t$, $p_t(i | k)$ denote the probability that an engagement of an item in interest $k$ is item $i$, and $p_t(k | u)$ denote the probability that user $u$'s next engagement is from interest $k$, we have: 
\begin{equation}
\label{eq:candgen}
p_t(i | u) = \sum_{k} p_t(i | k)p_t(k | u) = \sum_k \phi_{k,t}(i) \cdot \theta_{u}(k).  \\ 
\end{equation}

Our candidate retrieval strategy given user $u$ and number of candidates $m$ is to return the top $m$ items according to Equation~\ref{eq:candgen}. Note that computing $p_t(i,u)$ is $O(K)$ for a total of $O(I \cdot K)$ computations for each user, but that all candidates can be scored in parallel. Similarly, candidate sets for different users can be evaluated via Equation~\ref{eq:candgen} in parallel. In practice, however, we need to estimate top candidates quickly and find that enforcing and exploiting sparsity in $\theta_{u}$ and only evaluating candidate items $i$ such that $\phi_{k,t}(i)$ is not small where $\theta_{u}(k)$ is non-zero drastically improves performance (e.g. computing $p_t(i | u)$ is $O(||\alpha_u||_0)$ where $||\alpha_u||_0 << K$). These approximations and parallel retrieval qualities make retrieval via MiCRO feasible in a large scale setting even on a small compute instance.

\section{Experiments}
We now evaluate MiCRO empirically by turning to our motivating task -- candidate retrieval. We introduce the two social media datasets used for experimentation as well as the metrics and baselines by which we evaluate MiCRO. Finally, we present and summarize our empirical results, finding that MiCRO considerably outperforms these baselines across several metrics on both datasets.

\subsection{Datasets} \label{datasets}
\noindent
\textbf{TwitterFaveGraph\footnote{\url{https://huggingface.co/datasets/Twitter/TwitterFaveGraph}} (\texttt{fave}):} To accompany our research contribution, we release an open source dataset of user to Tweet engagement data which we refer to as \texttt{fave}. We curated this dataset by obtaining Tweet favorites from a set of users (available via API) and  subsampling this (user, \emph{favorites}, Tweet) graph. Each engagement is directed from user to Tweet. All users and Tweets are anonymized with no personally identifying information present in the data. Additionally, we bin the data into predetermined time chunks and assign them to ordinals. These ordinals are contiguous and respect time ordering, but do not provide any information on the exact time each engagement occurred. In total we have 283\emph{M} edges, 6.7\emph{M} user vertices, and 13\emph{M} item (Tweet) vertices. The maximum degree for users is 100 and the minimum degree for users is 1. The maximum degree for items (Tweets) is 280\emph{k} and the minimum degree for items is 5.
\ \\
\noindent
\textbf{TwitterFollowGraph\footnote{\url{https://huggingface.co/datasets/Twitter/TwitterFollowGraph}} (\texttt{follow}):} An open-source Twitter Follow Graph dataset~\cite{el2022knn} which we refer to as \texttt{follow}. This dataset is constructed by subsampling the (user, \emph{Follow}, user) graph and provides an ordinal time chunk indicating when the follow occurred. In total, the dataset has 261\emph{M} edges and 15.5\emph{M} vertices, with a max degree of 900\emph{k} and a min-degree of 5.

\subsection{Metrics}

We evaluate MICRO temporally against future held-out engagements on three standard candidate retrieval metrics: 

\begin{enumerate}
    \item \textbf{Recall@M}: Recall@M measures what proportion of relevant (ground truth) candidates were correctly retrieved by a model returning M candidates. High recall suggests many relevant items for a downstream ranking model to reorder while optimizing for precision.%
    
    \item \textbf{Mean Reciprocal Rank (MRR)}: Mean of the reciprocal ranks of the first relevant item in retrieved candidate sets for each user. In certain settings such as ranking push notifications we may only care about finding a single relevant item to send \citep{yue2022learning}. %
    
    \item \textbf{Normalized Discounted Cumulative Gain (NDCG)}: NDCG~\citep{10.1145/582415.582418} is a useful measure of a model's ability to not only retrieve relevant candidates, but also rank them comparatively higher than irrelevant candidates in the result set. %
\end{enumerate}

As our method is designed for trending and ephemeral content, we report our metrics both as overall averages over all (user, time chunk) data-points, as well as averages over users per time chunk. 
\subsection{Baselines}
\label{baseline}

We compare MiCRO against two baselines:

\begin{enumerate}
    \item encoding items as averages of embeddings of users who engaged with the item, followed by approximate nearest neighbor (ANN) cosine similarity retrieval
    \item a global popularity baseline where we take the highest engaged with items over some time period as a fixed recommendation for all users
\end{enumerate}

Baseline models were chosen for their (1) interpretability, (2) ability to efficiently model out-of-vocabulary ephemeral items, and (3) ability to model temporal trends. Our embedding ANN baseline that encodes items by averaging user vectors is analogous to the word embedding averaging used when representing sentences as averages of the words' embeddings\cite{goldberg2014word2vec, Socher2013RecursiveDM}. In sections below, we refer to (1) as \texttt{ANN} and (2) as \texttt{Popularity}.

\subsection{Experimental Setup}

In this section we will describe in detail any initial processing required to power MiCRO or either of the two baselines introduced in Section~\ref{baseline}. We will also cover the exact backtesting scheme and hyperparameter explorations we use to evaluate on the datasets introduced in Section~\ref{datasets}.

We start by creating a bipartite (user, item) interaction graph for each dataset. Edges in this dataset represent positive, relevant engagements between users and items (e.g. user \emph{follows} user, or user \emph{favorites} tweet).  We separate both datasets into \emph{train} and \emph{test} components. The \emph{train} data is a larger period of time where we can follow the techniques in Section~\ref{making_an_embedding} to converge a graph embedding for users and historical items and then use our spherical clustering technique describe in Section~\ref{clustering_and_projecting} to infer user interest mixtures. The \texttt{ANN} baseline makes use of the same underlying dense user embeddings that are trained as part of initializing MiCRO via item clusters. The \texttt{Popularity} baseline has no dependency on any embedding artifacts from \emph{train} data.

In the \emph{test} data, we use the user representations learned from \emph{train} and apply MiCRO in order induct new items and retrieve candidates for the next time chunk on a rolling basis - mimicking a real system that runs online or in batch over time. The two baselines are applied in the same way given their underlying methods of item induction and retrieval. That is, for some time \emph{t}, we will have some representation for users and items under MiCRO or the two baselines, and we will use these representations to retrieve candidates evaluated on ground truth from \emph{t+1}.

For \texttt{fave} we reserve the last 23 time chunks, out of 192 total in the data, for \emph{test} and for \texttt{follow} we first re-group the existing open source data into 25 coarser chunks, and then reserve the last 7 for \emph{test}. This leaves us with 34\emph{M} (user, time chunk) datapoints in the \emph{test} data for \texttt{fave} and 78\emph{M} (user, time chunk) datapoints in the \emph{test} data for \texttt{follow}.

For the initialization component of MiCRO and the dense user embeddings which power \texttt{ANN}, we train 128-dimensional embeddings for users and items on both datasets for 20 epochs. To cluster items into interests, we apply spherical \emph{k}-means for 25 epochs to cluster items based on their embedding vectors.

\ \\
\noindent
\textbf{Parameter Inference and Candidate Retrieval:} We utilize a collapsed Gibbs sampler to learn the latent parameters for MiCRO by optimizing the log-likelihood of the joint distribution between Users and Items at time \emph{t}. As a result, we learn user interest distributions $\theta_{u}$ and item-interest distributions $\phi_{k,t}$. However, these learned distributions are only useful in online retrieval settings if they can be used to make item recommendations for future engagements. We use our learned parameters from time \emph{t} to perform candidate retrieval at time \emph{t+1} and measure on our extrinsic candidate retrieval tasks of \emph{Recall}, \emph{MRR}, and \emph{NDCG}. By doing so we also explore the relationship between a MiCRO model optimizing log-likelihood on some historic but recent data, with a future-engagement retrieval objective. We expect the feasibility of success on this task to be dataset dependent based on many aspects such as the level of item ephemerality or temporality present in the data.

\ \\
\noindent
\textbf{Hypotheses:}

\begin{itemize}
    \item a global popularity baseline will not be personalized and retrieve many irrelevant items
    \item encoding via unimodal embedding aggregation and approximate nearest neighbor retrieval will retrieve more relevant candidates, but they will be highly intrasimilar and subsequently show lower performance on the recall task
    \item MiCRO will have the highest performance due to retrieving a diverse and relevant set of candidates pertaining to the diverse interests users may have
\end{itemize}

We would also like to have some understanding of how the interest count hyperparameter of MiCRO may affect the quality of the recommendation. As a followup, we include some exploration of retrieval at lower M in the Appendix.

\subsection{Results}

To evaluate the overall performance we computed our three metrics across all (user, time chunk) queries in the test data for both datasets. We present both the overall summarized data as well as a segmented time-series view split by time chunk in the test data. 

In the figures below we report our mean recall, MRR, and NDCG at a fixed $M = 100$ over time chunks in the test period. For each plot, we choose the number of latent interest clusters for MiCRO that performs best on the ``Overall Mean Recall@100'' benchmark.

\begin{figure*}[ht]
     \centering
     \begin{subfigure}[b]{0.315\textwidth}
         \centering
         \includegraphics[width=\textwidth]{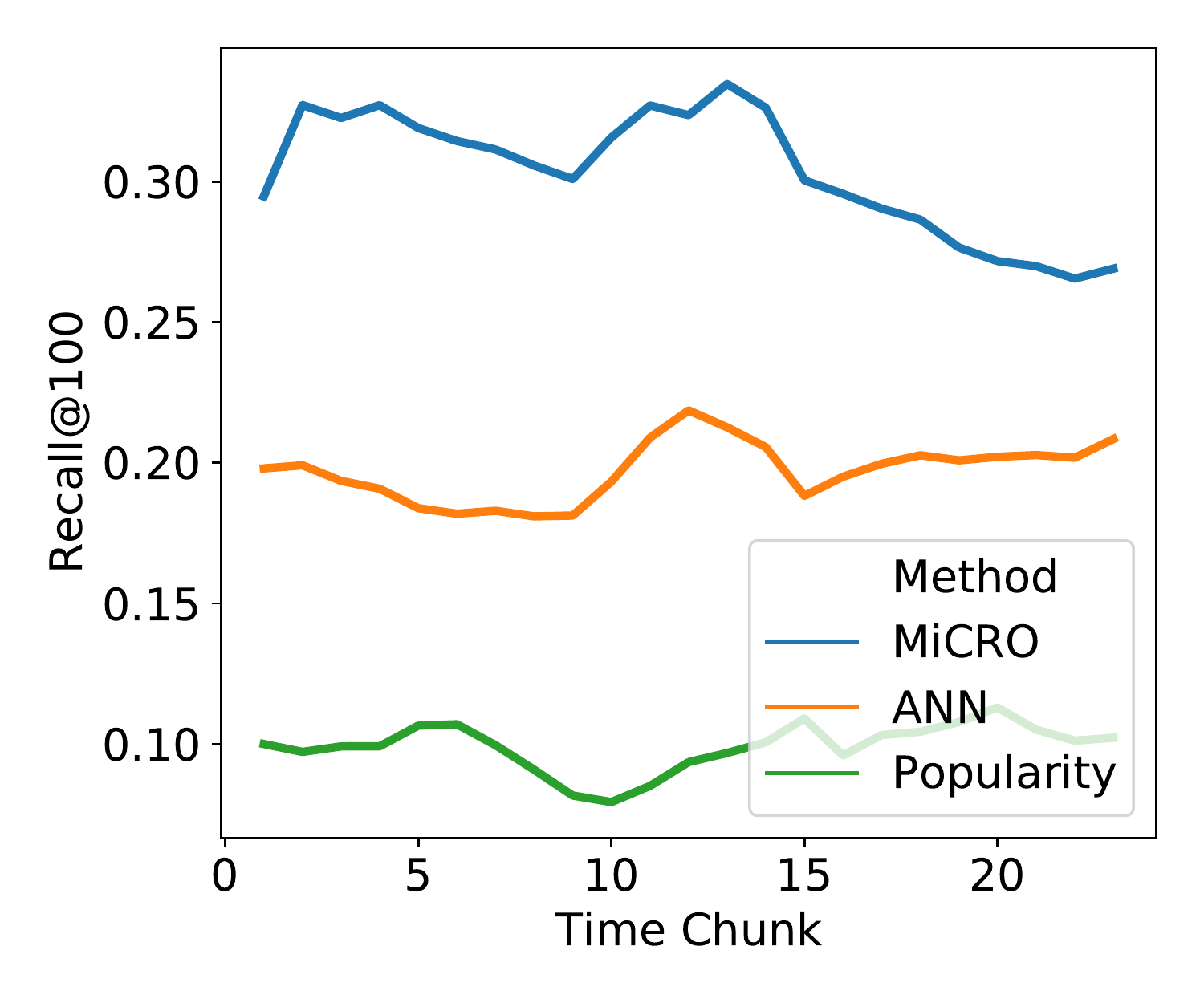}
         \caption{Recall over time.}
         \label{fig:fav_recall}
     \end{subfigure}
     \hfill
     \begin{subfigure}[b]{0.315\textwidth}
         \centering
         \includegraphics[width=\textwidth]{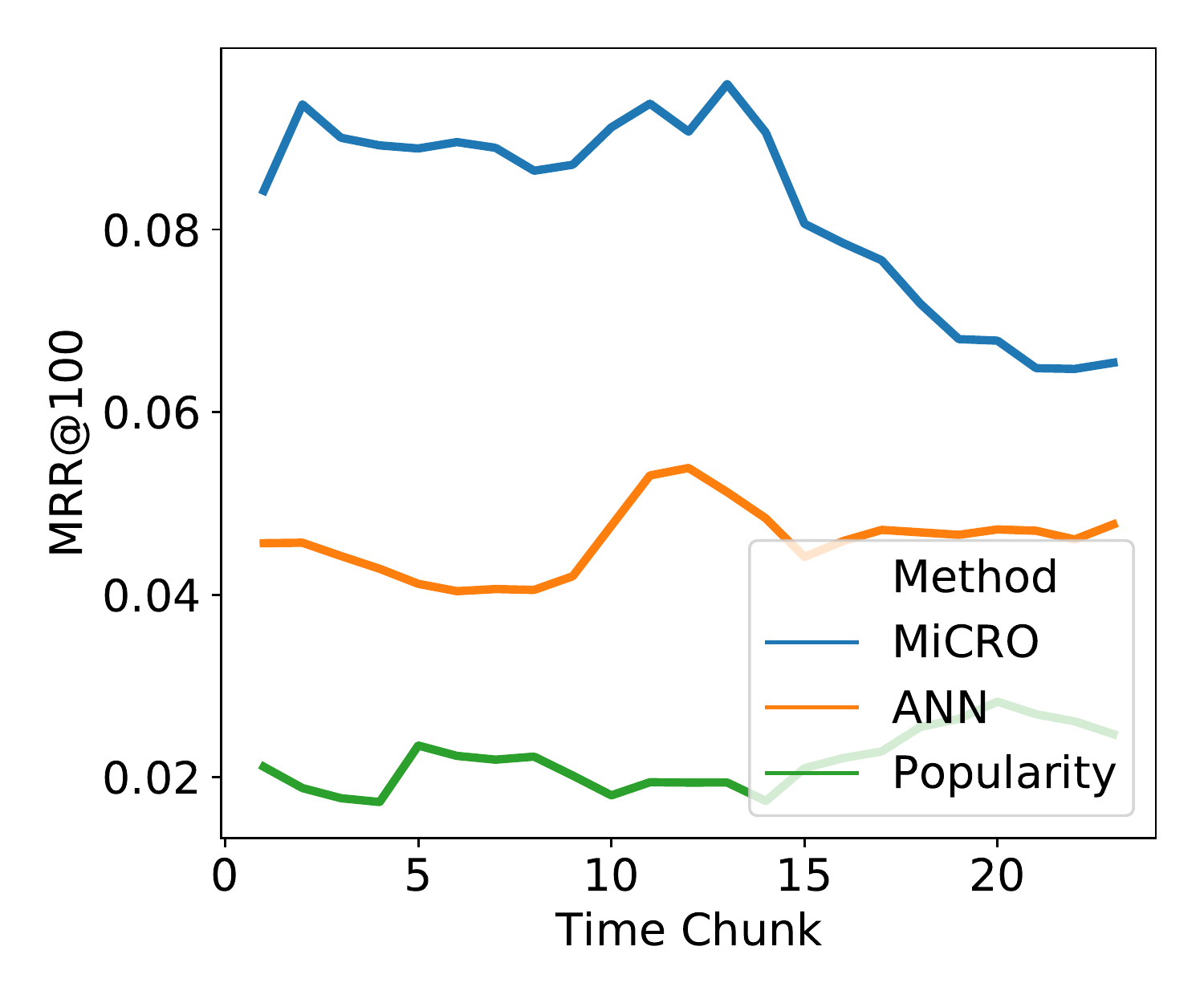}
         \caption{MRR over time.}
         \label{fig:fav_mrr}
     \end{subfigure}
     \hfill
     \begin{subfigure}[b]{0.302\textwidth}
         \centering
         \includegraphics[width=\textwidth]{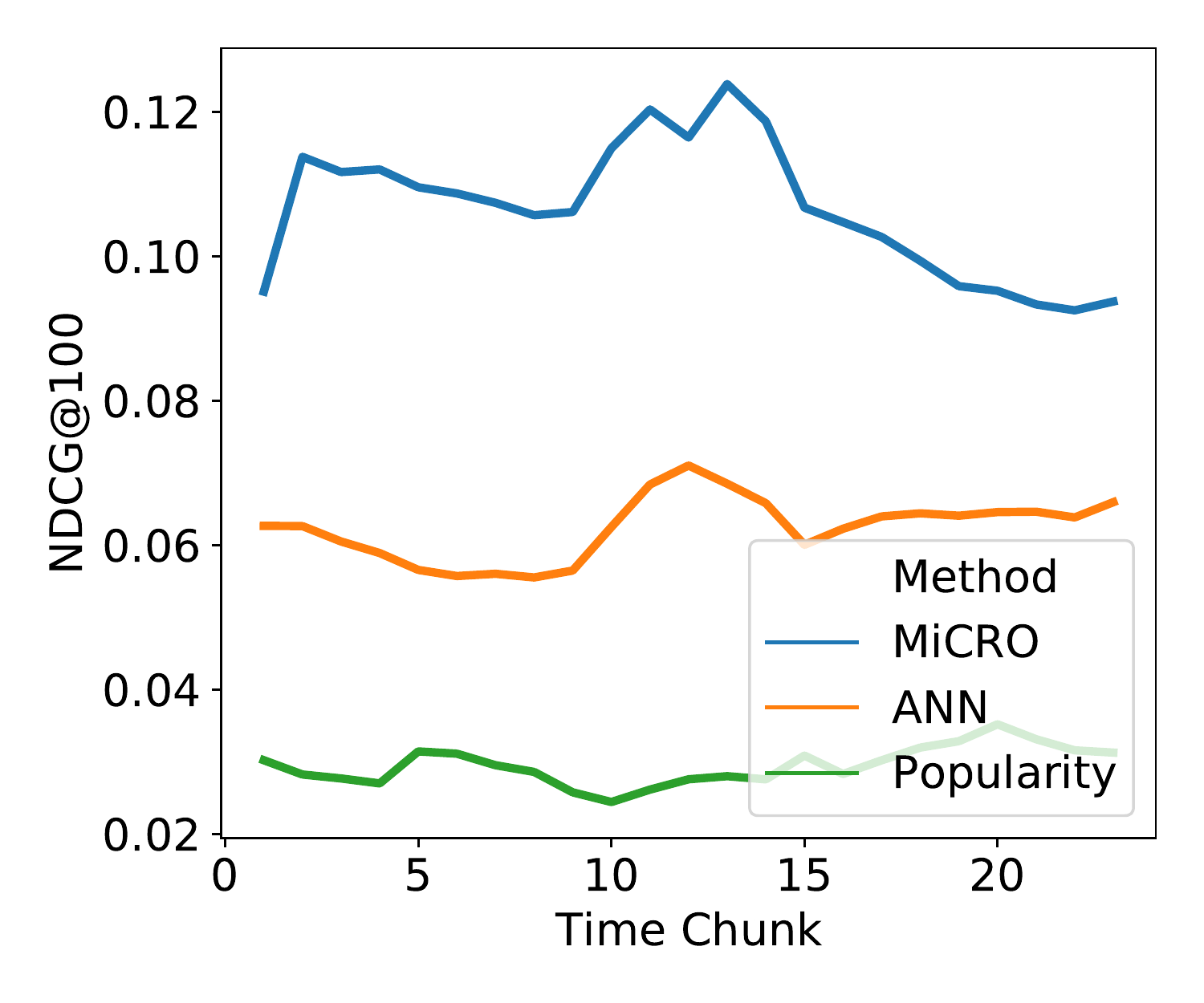}
         \caption{NDCG over time.}
         \label{fig:fav_ndcg}
     \end{subfigure}
        \caption{Mean Recall@100, MRR@100, and mean NDCG@100 for every time chunk in the \texttt{fave} test data for the highest performing MICRO and the baseline models. For MiCRO, the optimal number of interests was 5000.}
        \label{fig:user_fav}
\end{figure*}

\begin{figure*}[ht]
     \centering
     \begin{subfigure}[b]{0.315\textwidth}
         \centering
         \includegraphics[width=\textwidth]{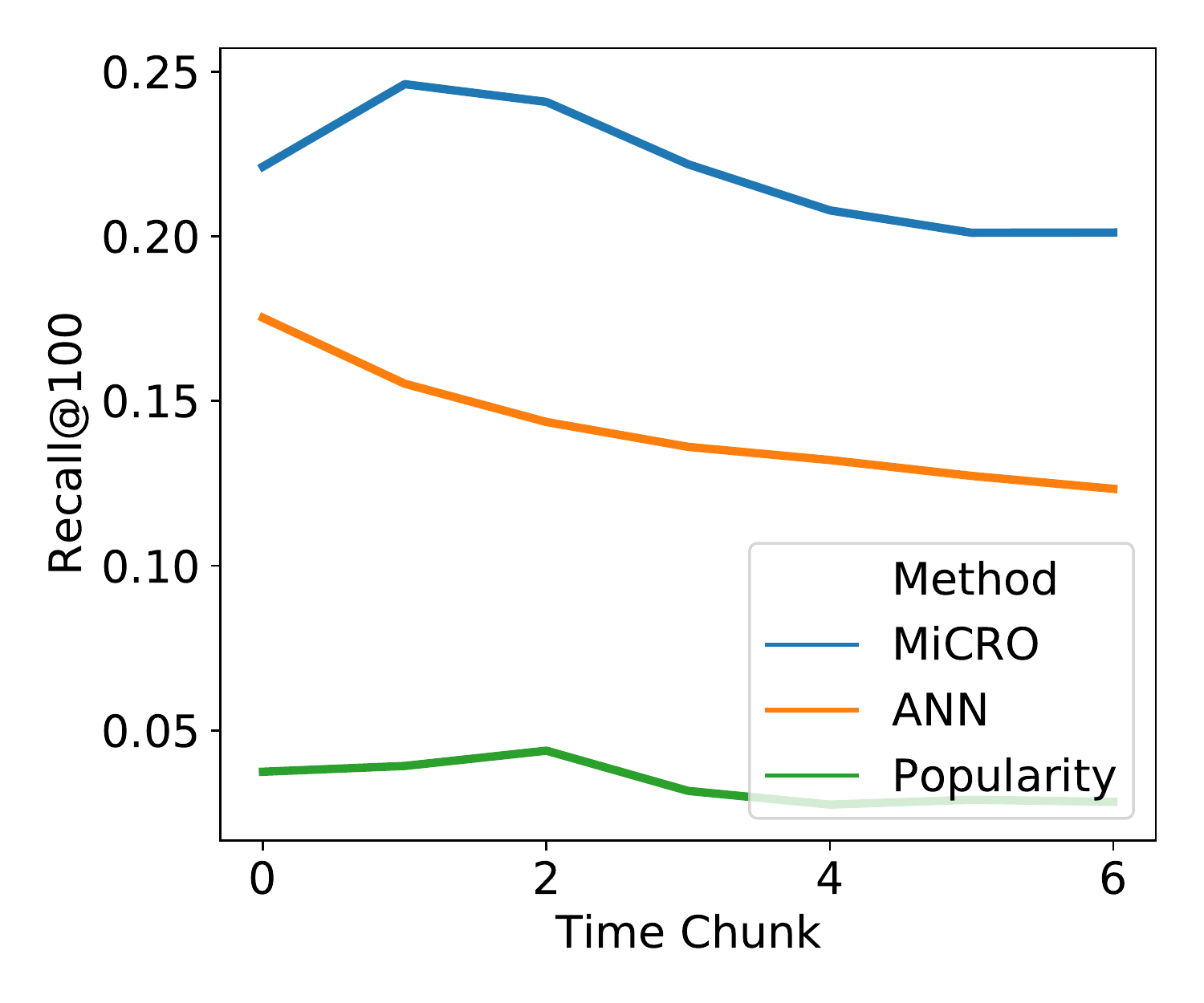}
         \caption{Recall over time.}
         \label{fig:follow_recall}
     \end{subfigure}
     \hfill
     \begin{subfigure}[b]{0.315\textwidth}
         \centering
         \includegraphics[width=\textwidth]{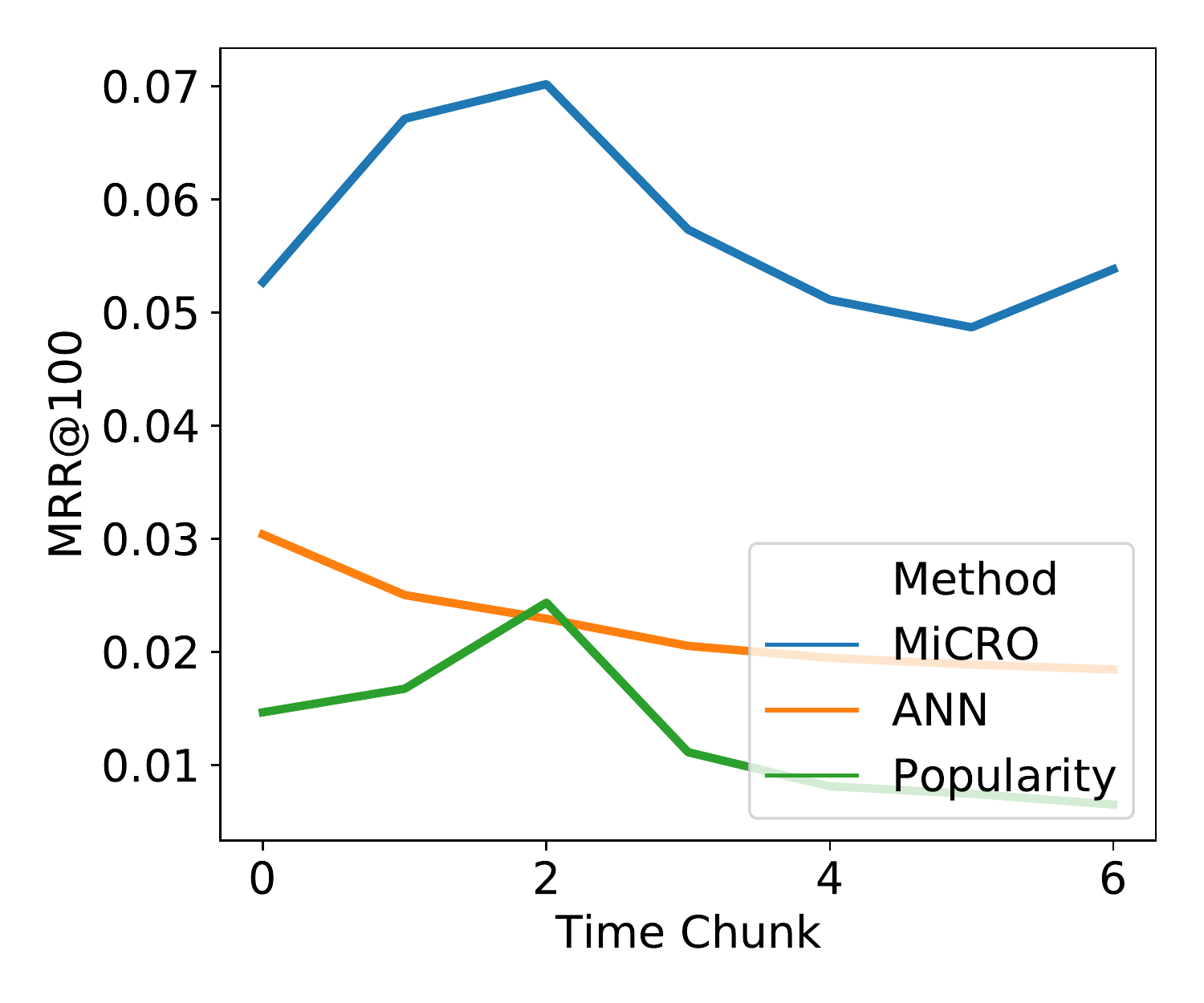}
         \caption{MRR over time.}
         \label{fig:follow_mrr}
     \end{subfigure}
     \hfill
     \begin{subfigure}[b]{0.315\textwidth}
         \centering
         \includegraphics[width=\textwidth]{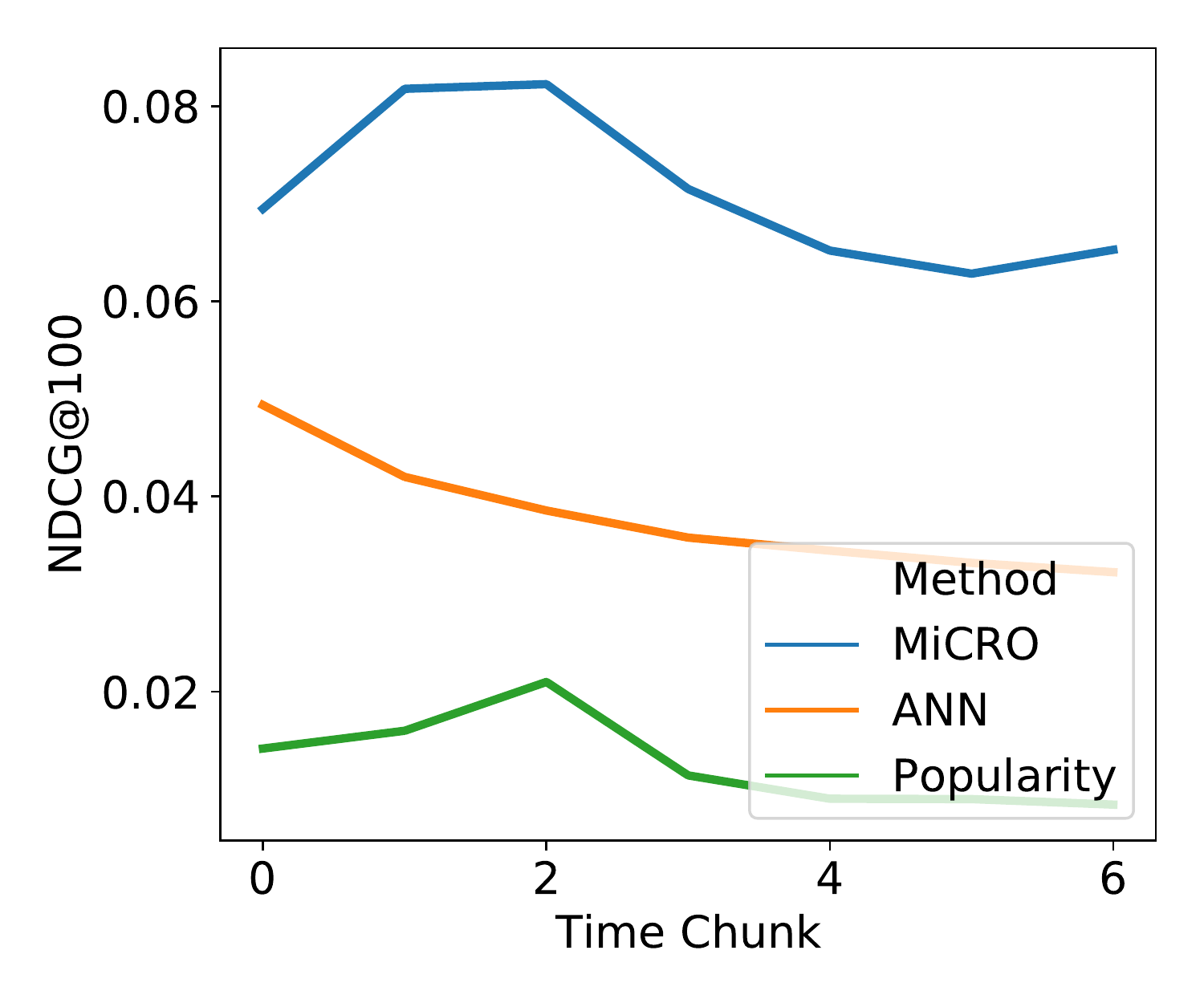}
         \caption{NDCG over time.}
         \label{fig:follow_ndcg}
     \end{subfigure}
        \caption{Mean Recall@100, MRR@100, and mean NDCG@100 for every time chunk in the \texttt{follow} test data for the highest performing MICRO and the baseline models. For MiCRO, the optimal number of interests was 5000.}
        \label{fig:user_follow}
\end{figure*}

Figure~\ref{fig:user_fav} compares MICRO with \texttt{ANN} and \texttt{Popularity} on the test data for \texttt{fave}. For each method, 100 candidates are retrieved for each user across each time chunk using representations from the previous time chunk and we compute Recall, NDCG, and MRR on the held-out ground truth engagements. We see that MICRO outperforms both \texttt{ANN} and \texttt{Popularity} on this data by a wide margin, showing for many time periods a nearly 50\% improvement in recall, and similar improvements over the baselines for MRR and NDCG. 

Figure~\ref{fig:user_follow} compares MiCRO with \texttt{ANN} and \texttt{Popularity} on \texttt{follow}. We again find that MICRO outperforms the ANN based methods at most time chunks by a considerable margin. However, on this dataset, \texttt{Popularity} baseline is closer in performance to \texttt{ANN} baseline and at one point outperforms. This is intuitive as when recommending users to follow, a user's popularity may be more important to consider versus any topical relevancy.

\begin{table}[h]
\label{tables}
\begin{subtable}[h]{0.45\textwidth}
\begin{tabular}{l | lll}
           & Recall@50         & MRR@50               & NDCG@50              \\
           \hline
MiCRO & $\textbf{0.161}$ & $\textbf{0.056}$ & $\textbf{0.060}$ \\
\texttt{ANN}  & 0.091          & 0.021          & 0.029          \\
\texttt{Popularity} & 0.023           & 0.013           & 0.011          
\end{tabular}
\caption{Overall performance on \texttt{follow} at M=50.}
\label{table:follow-overall-ir-metrics-50}
\end{subtable}

\begin{subtable}[h]{0.45\textwidth}
\begin{tabular}{l | lll}
           & Recall@100     & MRR@100        & NDCG@100       \\
           \hline
MiCRO & $\textbf{0.223}$ & $\textbf{0.057}$ & $\textbf{0.072}$ \\
\texttt{ANN}  & 0.143           & 0.023           & 0.038           \\
\texttt{Popularity} & 0.034           & 0.013           & 0.013          
\end{tabular}
\caption{Overall performance on \texttt{follow} at M=100.}
\label{table:follow-overall-ir-metrics-100}
\end{subtable}

\begin{subtable}[h]{0.45\textwidth}
\begin{tabular}{l | lll}
           & Recall@50    & MRR@50        & NDCG@50       \\
           \hline
MiCRO & $\textbf{0.226}$ & $\textbf{0.081}$ & $\textbf{0.087}$ \\
\texttt{ANN}  & 0.144          & 0.045         & 0.053            \\
\texttt{Popularity} & 0.068            & 0.022         & 0.024                 
\end{tabular}
\caption{Overall performance on \texttt{fave} at M=50.}
\label{table:fav-overall-ir-metrics-50}
\end{subtable}

\begin{subtable}[h]{0.45\textwidth}
\begin{tabular}{l | lll}
           & Recall@100     & MRR@100        & NDCG@100       \\
           \hline
MiCRO & $\textbf{0.306}$ & $\textbf{0.082}$ & $\textbf{0.102}$ \\
\texttt{ANN}  & 0.198        & 0.046       & 0.063       \\
\texttt{Popularity} & 0.101          & 0.022          & 0.030           
\end{tabular}
\caption{Overall performance on \texttt{fave} at M=100.}
\label{table:fav-overall-ir-metrics-100}
\end{subtable}
\caption{Tables of results for M=50 and M=100 on both datasets using 5k interests in MiCRO for the \texttt{fave} and \texttt{follow} datasets. 5k was the best performing number of interests on Recall@100 for both datasets.} 
\label{table:table-overall}
\end{table}

\begin{figure*}[h]
     \centering
     \begin{subfigure}[b]{0.45\textwidth}
         \centering
         \includegraphics[width=\textwidth]{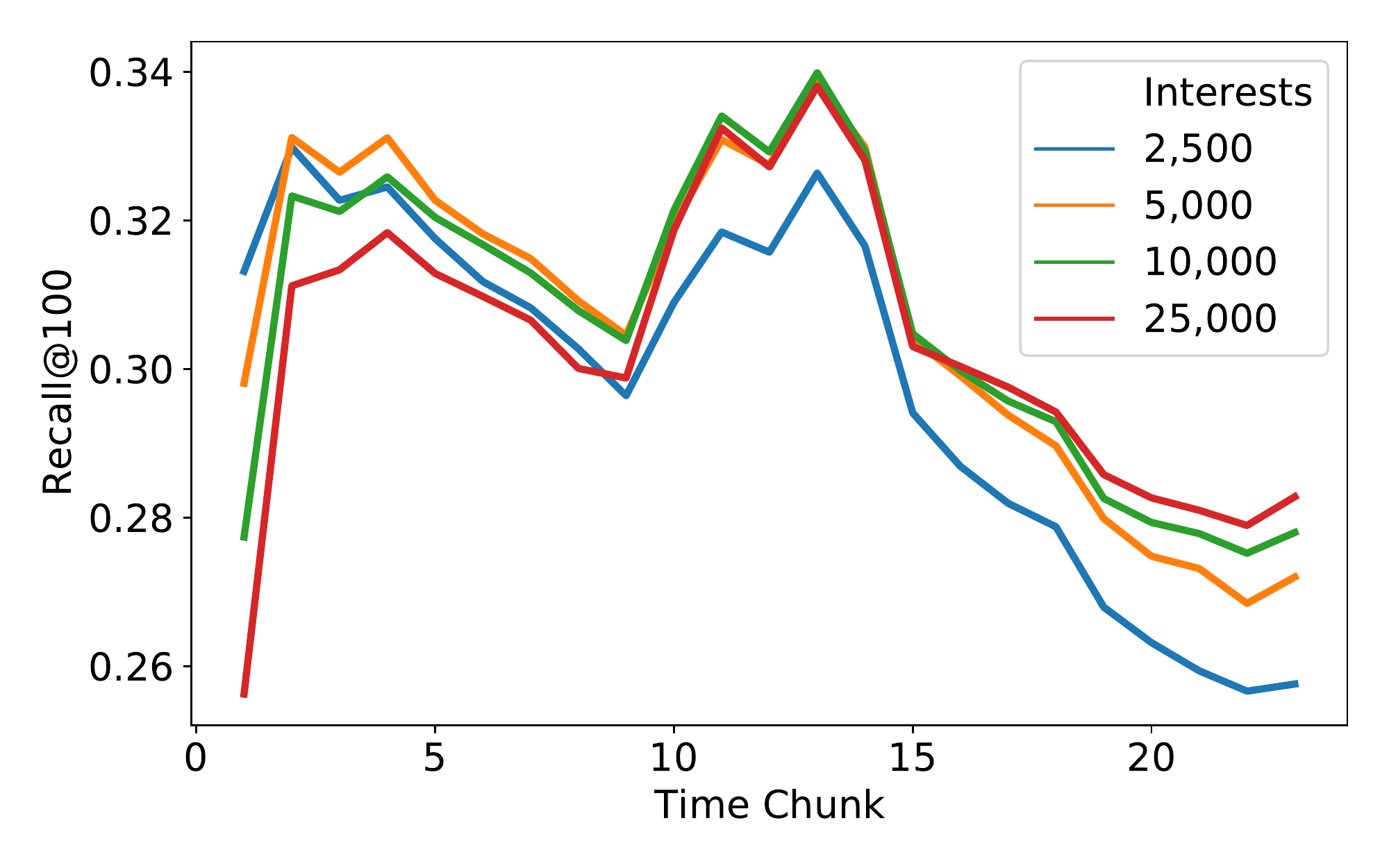}
        \caption{Varying number of interests in \texttt{fave}.}
         \label{fig:vary-num-interests-fav}
     \end{subfigure}
     \hfill
     \begin{subfigure}[b]{0.45\textwidth}
         \centering
         \includegraphics[width=\textwidth]{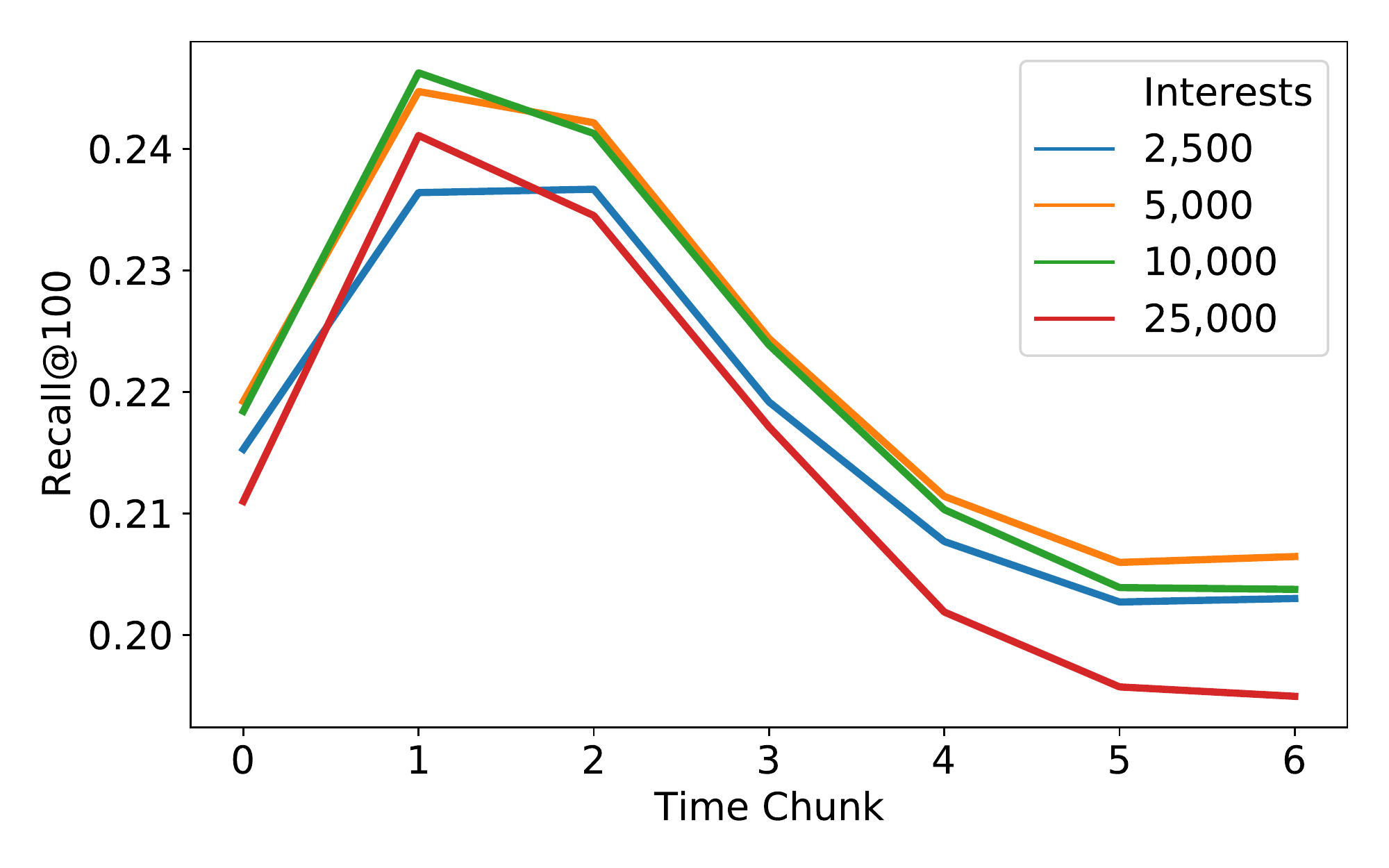}
        \caption{Varying number of interests in \texttt{follow}.}
         \label{fig:vary-num-interests-follow}
     \end{subfigure}
     \caption{MiCRO's Recall@100 for varying numbers of latent interests.}
     \label{fig:vary-interests}
\end{figure*}

In addition to evaluating our approach by looking at the average performance over each time chunk, we also consider aggregation at the user level over all time chunks. In Table~\ref{table:table-overall} we again find that MICRO outperforms the baselines significantly on our datasets at this more ``global" task at varying levels of $M$.

\ \\
\noindent
\textbf{Varying the number of interests:} Here we look at the impact of varying the number of interests we use in MiCRO. In Figure~\ref{fig:vary-num-interests-fav} we see that MiCRO's Recall@100 in the later Time Chunks generally improves as the number of latent interests is increased from 2500 to 25000, but in earlier windows fewer interests perform better. In Figure~\ref{fig:vary-num-interests-follow} we find that MiCRO's performance across different time chunks as a function of the number of latent interests is more homogeneous across time for the \texttt{follow} data, and see that 5000 interests is optimal for most time chunks, though narrowly behind 10000 on the earlier chunks. 2500 and 25000 are both comfortably below, suggesting that they might be too few and too many interests respectively to effectively represent this item space.

\subsection{Discussion}

In this paper we benchmark temporally adapted methods suited for trending and ephemeral item retrieval. We briefly discussed that MiCRO's expected efficacy in real settings, and the need for a framework such as MiCRO, depends on the level of temporality or ephemerality one expects in the data. Further work can investigate the concepts of temporality and ephemerality as descriptive properties of data by (1) providing additional analysis of MiCRO's performance over unimodal or multi-interest retrieval strategies that do not factor in temporal adaptations and (2) additionally considering datasets that might be expected to have low levels of temporal item relevance or ephemeral new item appeal.

We also believe it is useful to explore smoothing for Users who may have interest mixtures estimated from very few interactions. Smoothing via Users' neighbors may improve diversity, coverage, and the eventual online item representations that we build with MiCRO.

\section{Related Works}

\textbf{Sparse Candidate Retrieval:}
The earliest techniques in candidate retrieval were based on retrieving items represented by large sparse vectors (e.g., one-hot encodings). These methods have largely relied on scalable approaches to search for similar sparse vectors from large target collections~\cite{bayardo2007scaling,andoni2006near}. These approaches often apply innovative indexing and optimization strategies to scale similarity search. Other approaches such as SimClusters in ~\citet{10.1145/3394486.3403370} perform multiple queries on sparse interest clusters to obtain social media candidates.

\noindent
\textbf{Dense Candidate Retrieval:}
Deep neural approaches in recommender systems~\citep{covington2016deep} have proliferated the use of similarity-search candidate retrieval in dense embedding spaces. Dense candidate retrieval has been applied in contexts of both item-based retrieval~\citep{de2015semantics} and collaborative filtering approaches~\citep{zhang2016collaborative}. Some early approaches apply hashing-based techniques that map inputs and targets onto discrete partitions and selecting targets from the same partitions as inputs~\cite{weston2013label}. Later, with improvements in fast approximate nearest neighbor search\cite{shrivastava2014asymmetric,malkov2018efficient,johnson2019billion}, dense nearest-neighbor approaches have been applied for candidate retrieval.

\noindent
\textbf{Temporal Adaptation:}
The temporal distribution shift problem on social media such as Twitter has been studied in~\citet{preotiuc-pietro-cohn-2013-temporal,rijhwani-preotiuc-pietro-2020-temporally,luu-etal-2022-time,mireshghallah2022non}. These works explore domain adaptation techniques that re-retrain a  model to capture temporal change. Similarly to our approach, \citet{preotiuc-pietro-cohn-2013-temporal} suggest temporally-relevant items by keeping track of latest item frequencies (hashtag frequencies) as a prior to an adaptive naive Bayes classifier. In this work, we generalize beyond keeping track of frequent items by creating multiple distributions over items where each distribution corresponds to a semantically coherent interest and models preferences over items.

\noindent
\textbf{Topic Modeling:}
Probabalistic topic modeling such as LDA~\cite{blei2003latent} are a popular method of
of discovering abstract ``topics'' underlying a collection of documents. Within these topic models, a topic is typically modeled as a multinomial distribution over words, and frequent words related by a common theme are expected to have a large probability in a topic multinomial. Similarly, in MiCRO, ``interests'' are modeled as multinomials over items and popular items related by a common theme have high probability in an interest multinomial. Many approaches have extended traditional topic models to utilize link or engagement data. PHITS was introduced as an extension to PLSA to define a generative process for both a document's text and the other documents it links to~\cite{hofmann2001missing}. Under this model, words and documents are drawn from topic-specific discrete distributions. Later works extended this model to make it fully generative ~\cite{erosheva2004mixed}. Many later topic models such as the Relational Topic Model~\cite{chang2009relational} and Mixed Membership Stochastic Block Model~\cite{airoldi2009mixed} explicitly model links between two documents. 

\section{Conclusions}

In this paper we proposed MiCRO, a statistical framework for item induction and candidate retrieval designed to perform well in a setting with diverse user interests, rapid creation of out-of-vocabulary items, and temporal item appeal. We derived a Gibbs Sampler for propagating initial user embeddings through recent engagements to infer model parameters. As part of our derivation and parametrization we discuss several properties of MiCRO that make it tractable for large graphs. To test empirical performance of MiCRO, we applied our method to two large social media engagement datasets, one open-source dataset consisting of users following other users, and another corresponding to user-Tweet engagements that we open source along with this work. We found that MiCRO outperformed both an ANN baseline and a temporal popularity baseline on a bevy of standard retrieval metrics. Given the strong theoretical motivation for this method as well as its superior performance on our empirical data, MiCRO is a promising direction for building upon recent advances in candidate retrieval.

\bibliographystyle{ACM-Reference-Format}
\bibliography{main}

\appendix

\section{Appendix}
Here we present more detailed results comparing the baseline and MiCRO methods across different parameter configurations and information retrieval tasks on the \texttt{fav} and \texttt{follow} datasets.

\begin{figure*}[ht]
     \centering
     \begin{subfigure}[b]{0.315\textwidth}
         \centering
         \includegraphics[width=\textwidth]{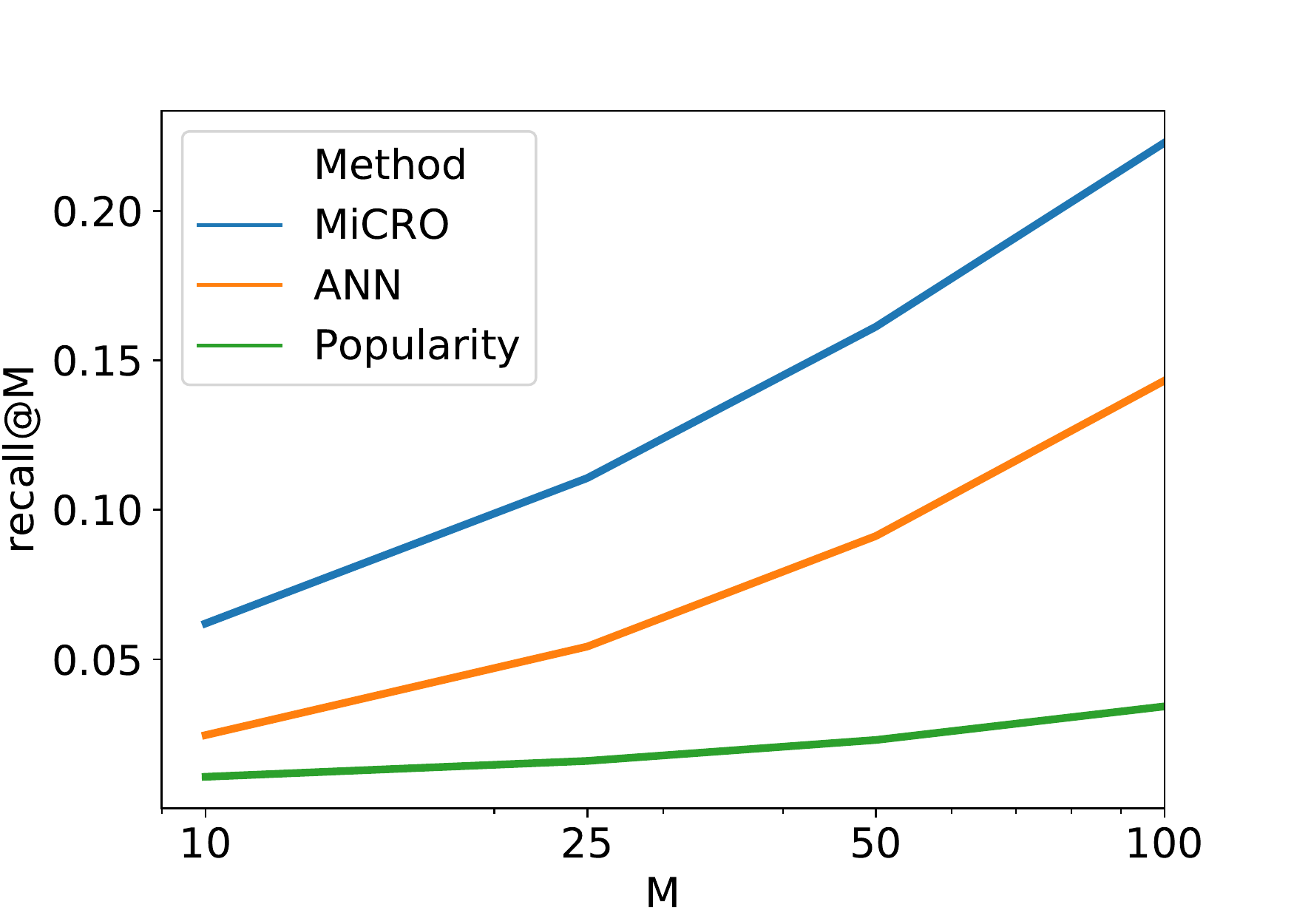}
         \caption{Recall@$M$.}
         \label{fig:vary_M_recall_wtf}
     \end{subfigure}
     \hfill
     \begin{subfigure}[b]{0.315\textwidth}
         \centering
         \includegraphics[width=\textwidth]{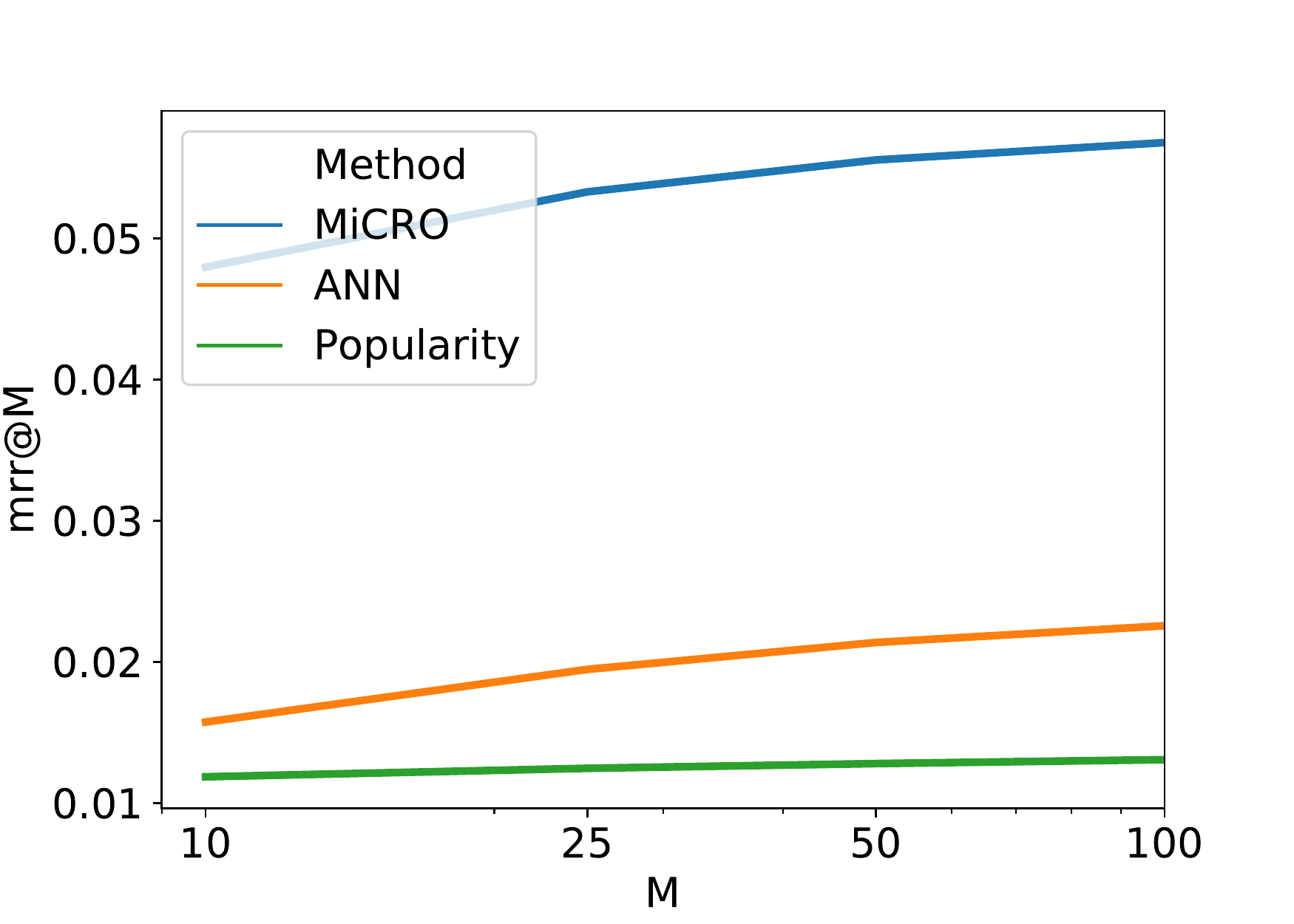}
         \caption{MRR@$M$.}
         \label{fig:vary_M_mrr_wtf}
     \end{subfigure}
     \hfill
     \begin{subfigure}[b]{0.315\textwidth}
         \centering
         \includegraphics[width=\textwidth]{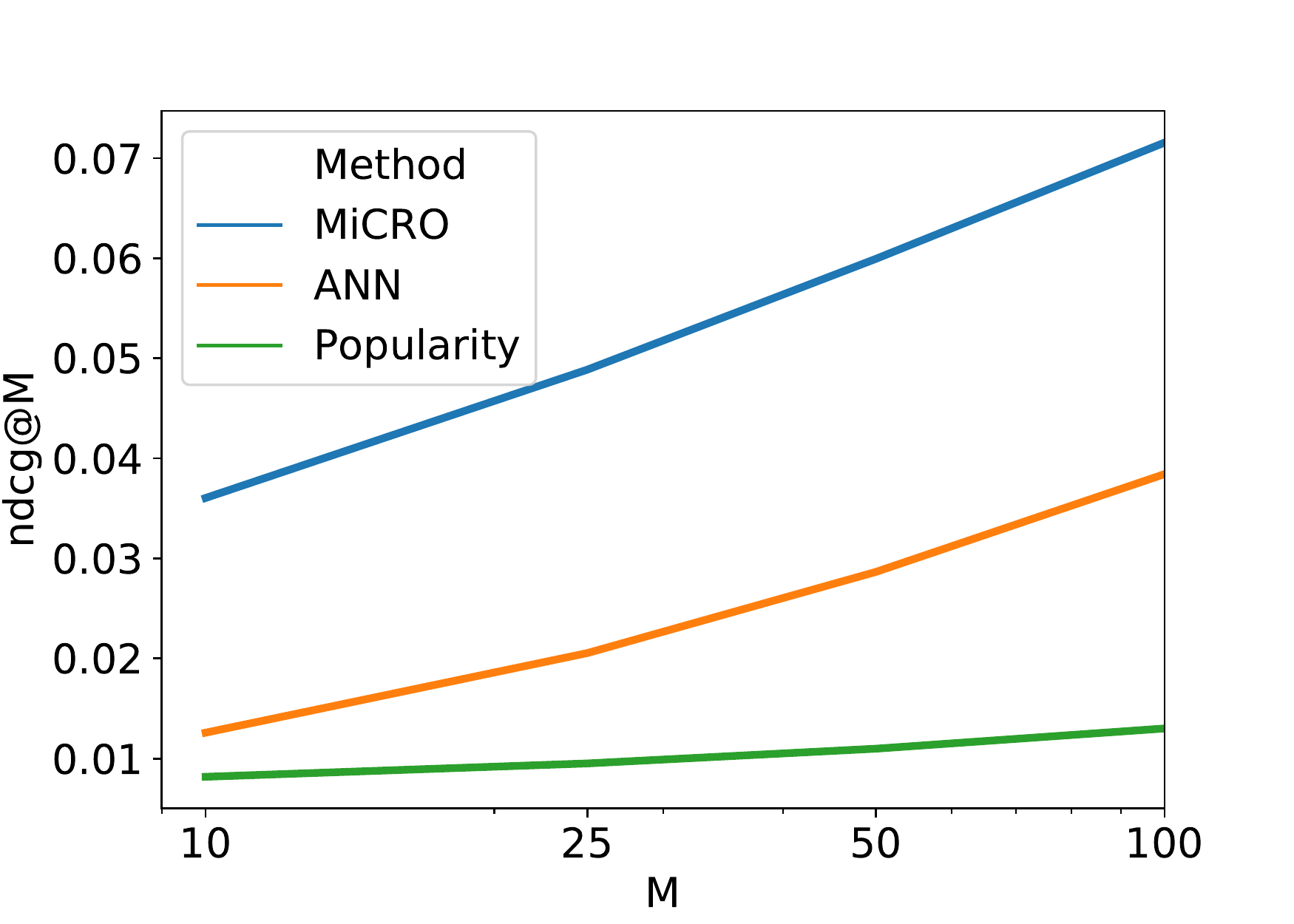}
         \caption{NDCG@$M$.}
         \label{fig:vary_M_ndcg_wtf}
     \end{subfigure}
        \caption{Mean Recall@M, MRR@M, and NDCG@M computed while varying $M$ on the \texttt{follow} test data.}
        \label{fig:vary_M_wtf}
\end{figure*}

\begin{figure*}[ht]
     \centering
     \begin{subfigure}[b]{0.315\textwidth}
         \centering
         \includegraphics[width=\textwidth]{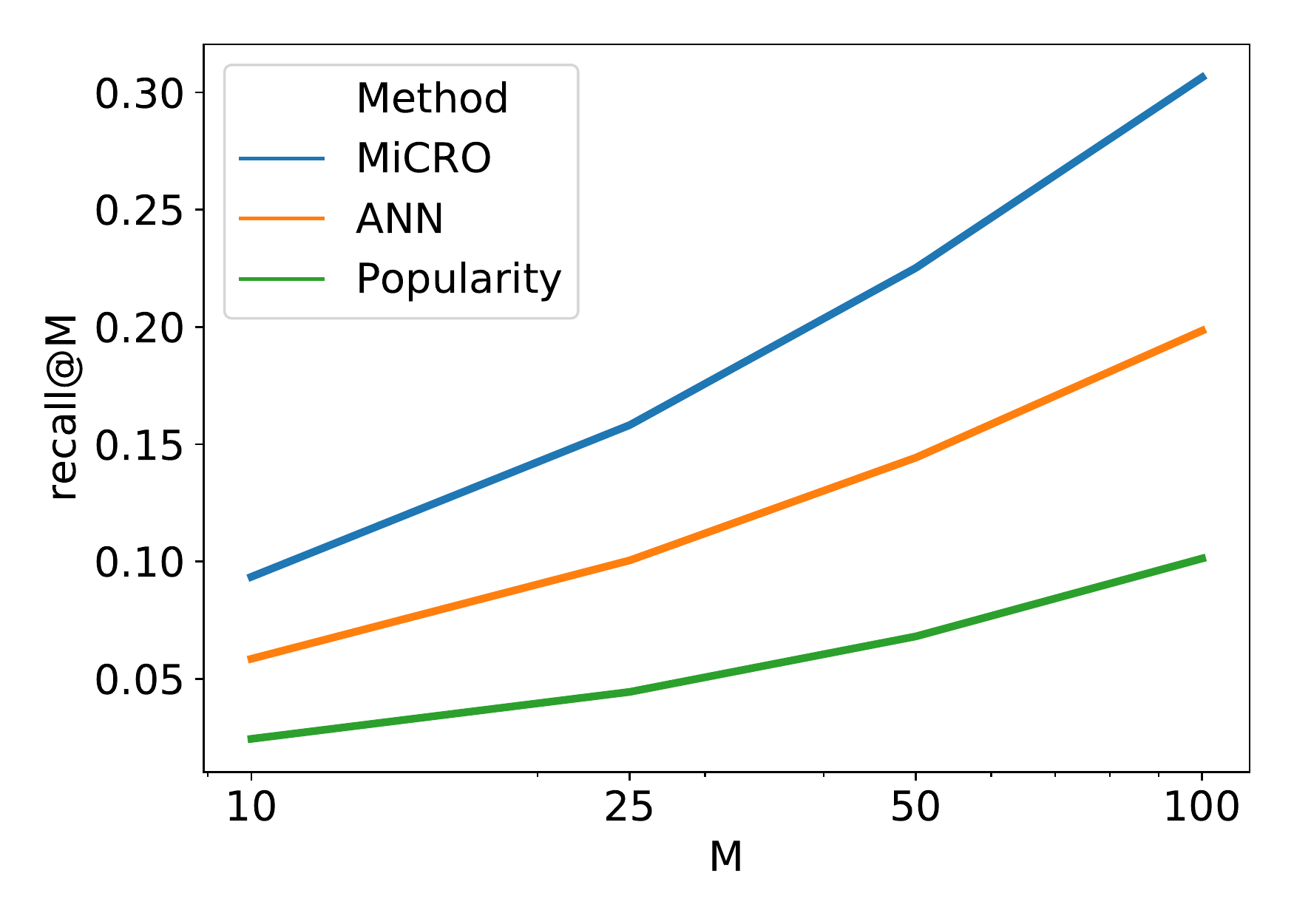}
         \caption{Recall@$M$.}
         \label{fig:vary_M_recall_fav}
     \end{subfigure}
     \hfill
     \begin{subfigure}[b]{0.315\textwidth}
         \centering
         \includegraphics[width=\textwidth]{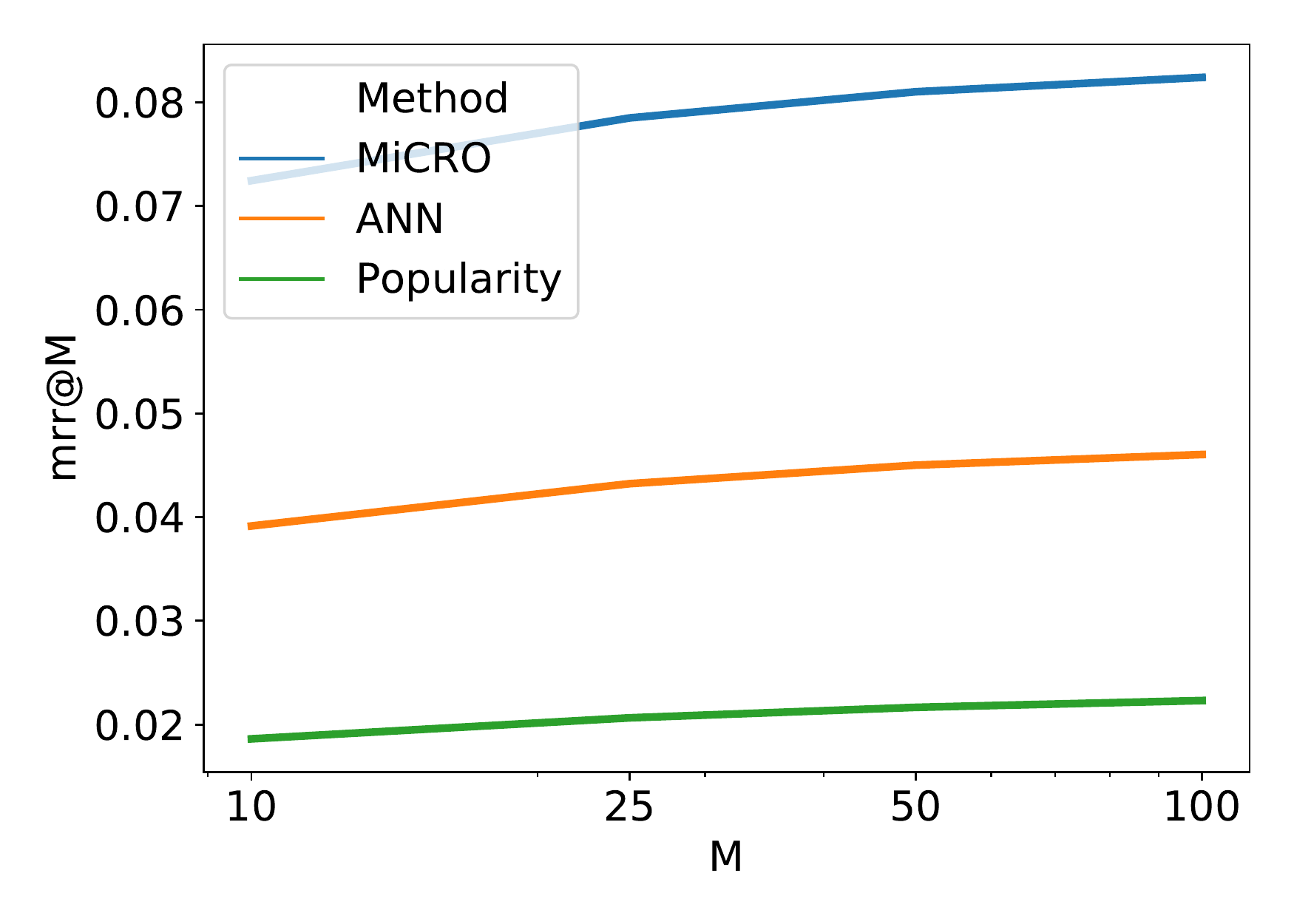}
         \caption{MRR@$M$.}
         \label{fig:vary_M_mrr_fav}
     \end{subfigure}
     \hfill
     \begin{subfigure}[b]{0.315\textwidth}
         \centering
         \includegraphics[width=\textwidth]{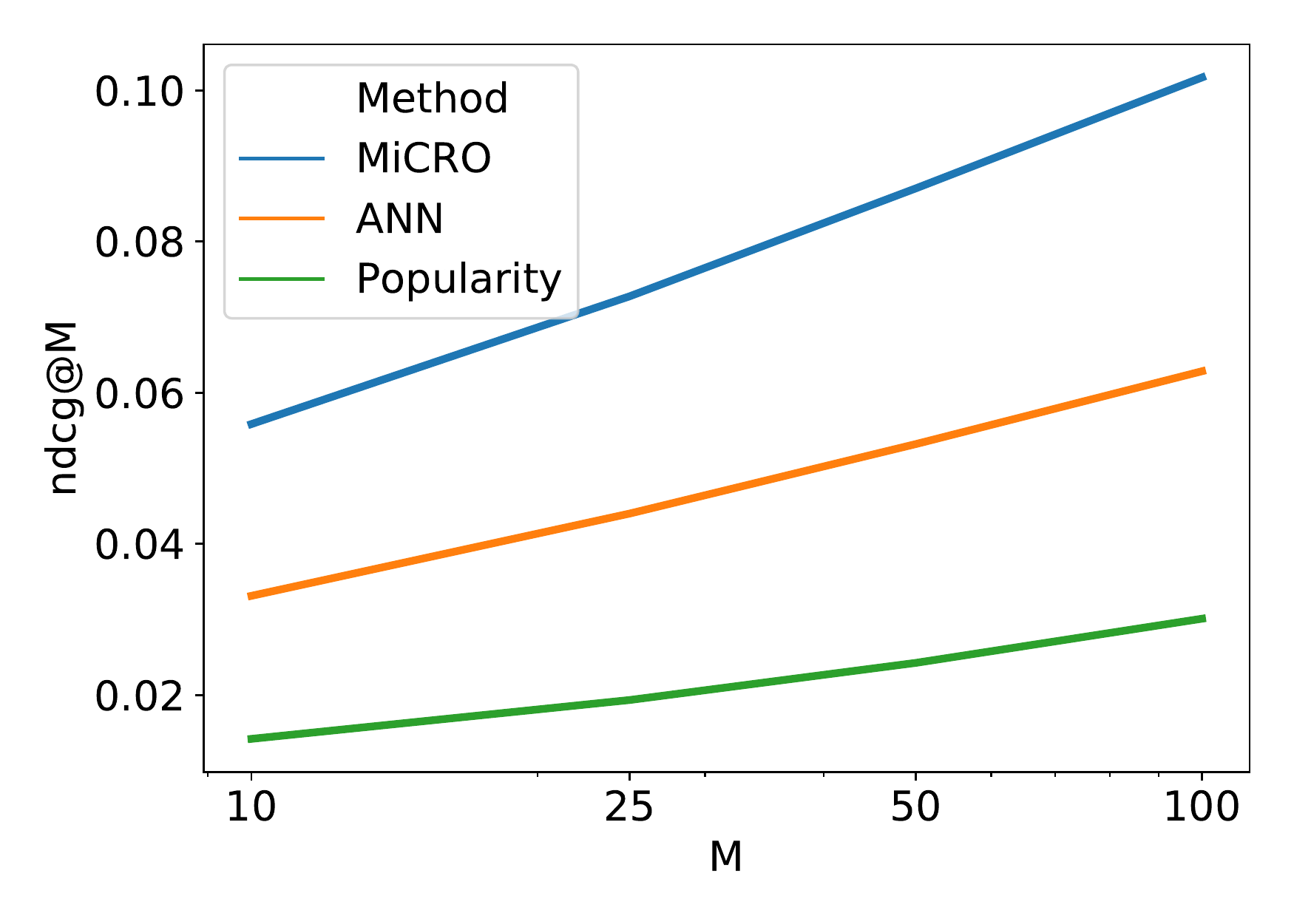}
         \caption{NDCG@$M$.}
         \label{fig:vary_M_ndcg_fav}
     \end{subfigure}
        \caption{Mean Recall@M, MRR@M, and NDCG@M computed while varying $M$ on the \texttt{fav} test data.}
        \label{fig:vary_M_fav}
\end{figure*}

While we focus on cases where the number of candidates, $M$, is 50 or 100 in the main paper, smaller $M$ narrow the absolute gaps between the different methods' performance on different metrics, but the relative performance still clearly shows MiCRO strongly outperforming \texttt{ANN} and \texttt{Popularity} across all metrics and values of $M$. 

In Figure~\ref{fig:vary_M_fav}, we compare Recall@M, MRR@M, and NDCG@M for the \texttt{fav} dataset of the same methods from Figure~\ref{fig:user_fav} but with $M=10,25,$ and $50$ as well as $M=100$. In Figure~\ref{fig:vary_M_recall_fav} we see that the gap between MiCRO and \texttt{ANN} on Recall@M appears for small $M$ and seems to stay steady over time, while both MiCRO and \texttt{ANN} continue to improve relative to \texttt{Popularity} as $M$ grows. We see a similar trend for MRR and NDCG in Figures~\ref{fig:vary_M_mrr_fav} and \ref{fig:vary_M_ndcg_fav} respectively - MiCRO again shows a large improvement over\texttt{ANN} on both MRR@M and NDCG@M for small $M$. This improvement stays roughly constant in $M$ while both methods pull away from \texttt{Popularity} baseline as $M$ increases.

We plot similar comparisons for the \texttt{follow} dataset in Figures~\ref{fig:vary_M_wtf}, again showing Recall@M, MRR@M, and NDCG@M for the baselines and MiCRO parametrization selected by Recall@100 where $M=10,25,50,100$. In Figure~\ref{fig:vary_M_recall_wtf} we see that for Recall@M both the gap between MiCRO and \texttt{ANN} and the gap between \texttt{ANN} baseline and \texttt{Popularity} grow as $M$ grows. In Figure~\ref{fig:vary_M_mrr_wtf} we see that for MRR@M both the gap between MiCRO and \texttt{ANN} and the gap between \texttt{ANN} and \texttt{Popularity} grow for small $M$ and seem to stabilize for larger $M$. In Figure~\ref{fig:vary_M_ndcg_wtf} we see that for NDCG@M both the gap between MiCRO and \texttt{ANN} and the gap between \texttt{ANN} and \texttt{Popularity} baseline grow as $M$ grows.

\section{Use of Data}

We study candidate retrieval on large scale temporal graphs with a focus on methods over applications. MiCRO can, in practice, be applied in industrial recommender systems. To the extent that our experimental data comes from existing systems or organizations, we do not make any claims as to how those systems generate recommendations in practice.

\end{document}